\documentclass[a4paper,fleqn,usenatbib]{mnras}

\usepackage{newtxtext,newtxmath}

\usepackage[T1]{fontenc}
\usepackage{ae,aecompl}

\usepackage{graphicx}	
\usepackage{amsmath}	
\usepackage{amssymb}	
\usepackage{etoolbox}
\makeatletter
\patchcmd\@combinedblfloats{\box\@outputbox}{\unvbox\@outputbox}{}{%
   \errmessage{\noexpand\@combinedblfloats could not be patched}%
}%
 \makeatother

\title[Impacts of the New C-Fusion Cross
Sections on SNe Ia]{Impacts of the New Carbon Fusion Cross
Sections on Type Ia Supernovae}

\author[K. Mori et al.]{
Kanji Mori$^{1,2,3}$,\thanks{E-mail: kanji.mori@nao.ac.jp}
Michael A. Famiano$^{4,3}$,
Toshitaka Kajino$^{1,2,3}$,
\newauthor \;Motohiko Kusakabe$^{3,1}$, and Xiaodong Tang$^{5}$
\\
$^{1}$National Astronomical Observatory of Japan, 2-21-1 Osawa, Mitaka, Tokyo,
181-8588 Japan\\
$^{2}$Graduate School of Science, The University of Tokyo, 7-3-1 Hongo, Bunkyo-ku,
Tokyo, 113-0033 Japan\\
$^{3}$School of Physics and Nuclear Energy Engineering, Beihang University, 37 Xueyuan Road, Haidian-qu, Beijing
100083, China\\
$^{4}$Department of Physics, Western Michigan University, Kalamazoo, Michigan 49008
USA\\
$^5$Institute of Modern Physics, Chinese Academy of Science, Lanzhou, Gansu 730000, China
}

\date{Accepted XXX. Received YYY; in original form ZZZ}

\pubyear{2018}

\begin{document}
\label{firstpage}
\pagerange{\pageref{firstpage}--\pageref{lastpage}}
\maketitle

\begin{abstract}
Type Ia supernovae (SNe Ia) are thought to be thermonuclear explosion of white dwarfs (WDs). Their progenitors are not well understood. One popular scenario is the double-degenerate (DD) scenario, which attributes SNe Ia to WD-WD binary mergers. The fates of the WD mergers depend on the rate of $^{12}$C+$^{12}$C reaction. Recently, the $^{12}$C+$^{12}$C cross sections {have been measured and the} analysis of the data using the Trojan Horse Method {suggested} that the astrophysical reaction rate is larger than conventional rates at  astrophysical temperatures {due to possible resonances}. {The resonance contribution results in a decrease of the carbon burning ignition temperature. Therefore accretion induced collapse occurs more easily and increases the birthrate of Galactic neutron stars with} the contribution of the DD scenario to the SNe Ia rate becoming even smaller. 
\end{abstract}



\section{Introduction}
The carbon fusion reactions $^{12}$C($^{12}$C, $\alpha$)$^{20}$Ne ($Q_\alpha=4.6$ MeV) and $^{12}$C($^{12}$C, $p$)$^{23}$Na ($Q_p=2.2$ MeV) play the important roles in stellar evolution and explosive phenomena in the Universe \citep[e.g.][]{iliadis,clayton}. The Gamow peak of these reactions is $1.5$ MeV at a temperature of $5\times10^8$ K, which is typical in astrophysical environments. Experimentalists have pursued these reaction cross sections in the sub-Coulomb energy for many years, but the cross section below 2.1 MeV has not been reported with direct methods \citep{exp1,exp2,exp3,exp4,exp5,becker,aguilera,barron,spillane,zickefoose}. 

Many resonances which can be interpreted as molecular resonances \citep{imanishi,chiba} are suggested experimentally using indirect methods \citep{kawabata} above the $^{12}$C+$^{12}$C threshold energy of the $^{24}$Mg$^\ast$ compound system. \citet{cooper} developed the idea of a low energy resonance near the Gamow peak. They assumed a resonance at $E=1.5$ MeV which does not contradict the available cross section data and applied it to X-ray superbursts {the ignition mechanism of which} is still unclear. \citet{bravo} and \citet{bennett} considered similar low energy resonances in order to apply them to an accreting white dwarf (WD) and evolution of massive stars, respectively. These assumed resonances lead to significantly enhanced reaction rates compared with the standard non-resonant rate given by \citet{CF88} (hereafter CF88).

{The resonance parameters proposed in the previous works were chosen so that resultant cross sections do not exceed the available cross section data at $E\approx2.1$ MeV.} However, it is unclear whether the chosen resonances are practically possible from the point of view of nuclear physics. In particular, the partial widths should be smaller than the Wigner limit \citep{teichmann,clayton}.

Recently, the cross sections were measured  for $E=0.8$ MeV to 2.7 MeV using the Trojan Horse Method \citep{tumino18}. Low-energy resonances which enhance the reaction rate by more than 25 times at $T\approx 5\times 10^8$ K compared with CF88 were found. These can have significant impacts on a wide range of astrophysics.

In this paper, we focus on the impact of the enhanced reaction rates on WD-WD binary mergers. It is suggested that they are progenitors not only of type Ia supernovae (SNe Ia) \citep{iben,webbink} but also of short $\gamma$-ray bursts \citep[e.g.][]{levan} and fast radio bursts \citep{kashiyama}, though their {evolutions and final fates} are still under debate. The reaction rate of carbon fusion is the most important input in this system because the reaction chain begins from $^{12}$C+$^{12}$C due to its small electric charge among all stable nuclei in the burning layers.

\citet{keane} estimated birthrates of Galactic neutron stars (NSs) and pointed out that they exceed the Galactic core-collapse supernova (CCSN) rate. One possible solution for this birthrate problem is the existence of sources of NSs other than CCSNe. It has been pointed out that the carbon burning flame in a WD-WD merger can turn a carbon-oxygen (CO) WD into an oxygen-neon-magnesium (ONeMg) WD \citep{nomoto91}. The ONeMg WD cannot support its mass and collapses into a NS. This evolutionary path enhances the NS birthrates, therefore it is worthwhile to estimate an event rate for this path. The condition for carbon burning to ignite and {evolve} a CO WD into a NS depends on the $^{12}$C+$^{12}$C reaction rate.

Here, we apply the new resonant carbon fusion reaction rate to the WD-WD mergers and discuss their fate in the context of the birthrate problem of Galactic NSs.

\section{Constraint on the Resonances}
{The partial decay width $\Gamma_\mathrm{C}$ for the $^{12}$C+$^{12}$C channel at astrophysically low energies $E<2$ MeV  is too far to reach both experimentally and theoretically because it is expected to be extremely small due to the Coulomb barrier between two carbons. Partial widths larger than the Wigner limit are practically impossible in terms of nuclear structure \citep[e.g.][]{clayton}, therefore we adopt it as {a conservative upper limit} on the width:
\begin{align}
\Gamma_\mathrm{C}(E_\mathrm{R})=2\gamma^2P_\mathrm{C}(E_\mathrm{R})=2\gamma_\mathrm{W}^2P_\mathrm{C}(E_\mathrm{R})\theta^2<2\gamma_\mathrm{W}^2P_\mathrm{C}(E_\mathrm{R}),
\label{upper}
\end{align}
where $P_\mathrm{C}$ is the Coulomb penetration factor, $E_\mathrm{R}$ is the resonance energy, $\gamma^2$ is the reduced width, and $\gamma^2_\mathrm{W}=3\hbar^2/2\mu a^2$ is the Wigner limit. Here, $\mu$ is the reduced mass and $a$ is the channcel radius. The partial widths are often parametrized by the dimensionless reduced width $\theta^2=\gamma^2/\gamma_\mathrm{W}^2$, i.e. $\theta^2=1$ is the Wigner limit.}

\begin{figure}
\centering
\includegraphics[width=8cm]{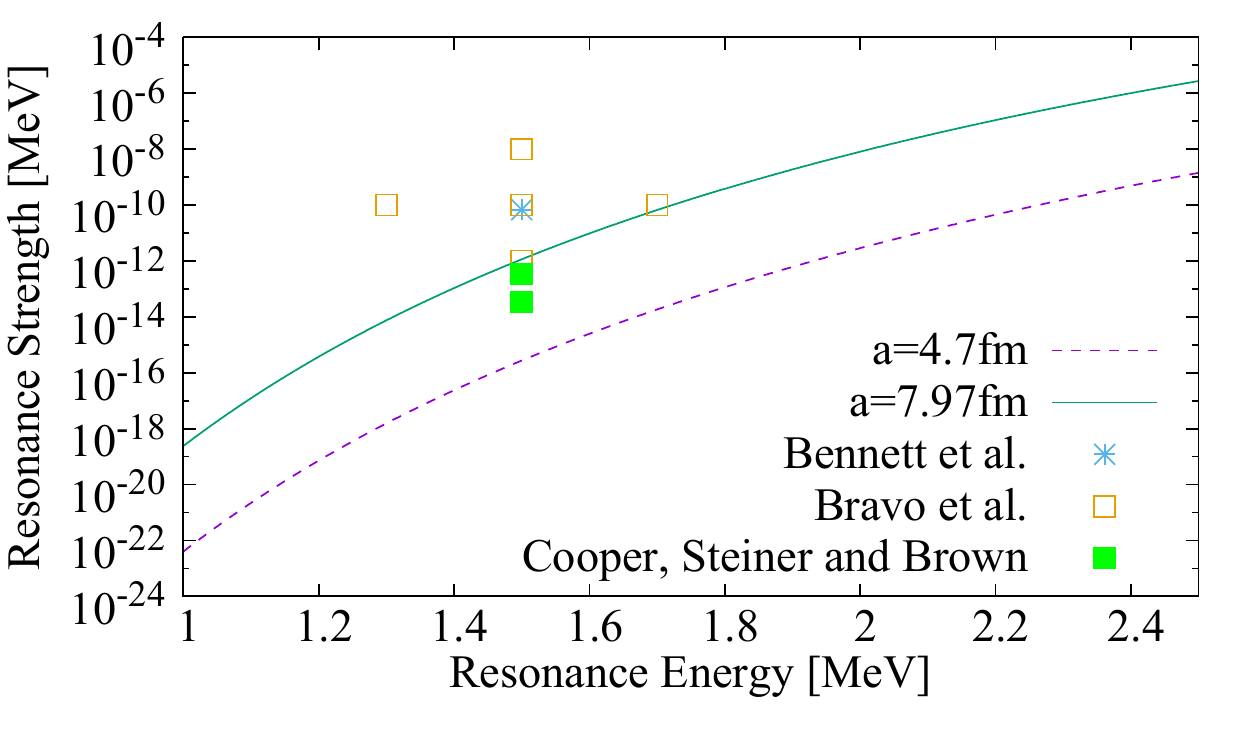}
\caption{The resonance strength with $\theta^2=1$ (Wigner limit) as a function of the resonance energy $E_\mathrm{R}$. The broken and solid lines are for the cases of $a=4.70$ fm and 7.97 fm, respectively \citep{kanungo,yakovlev}. The total angular momentum of the resonance is assumed to be $J=0$. The asterisk, and open and closed squares show the resonance parameters adopted in the previous works \citep{bennett,bravo,cooper}.}
\label{fig:strength}
\end{figure}
{Fig. \ref{fig:strength} shows the upper limit of the resonance strength deduced from Eq. (\ref{upper}) and resonance parameters are the same as those adopted in the previous works \citep{bennett,bravo,cooper}. The channel radius $a$ of the $^{12}$C$+^{12}$C channel is subject to a large uncertainty. The Coulomb penetration factor and the Wigner limit are highly dependent on this $a$ value. \citet{kanungo} derived a matter radius of 2.35 fm from the measurements of charge exchange reaction cross section, thus the channel radius for the $^{12}$C+$^{12}$C fusion reaction is estimated to be simply double of a matter radius of single carbon nucleus, $a=4.70$ fm. A model fitting of the astrophysical $S$-factor by \citet{yakovlev} also suggests a channel radius as large as $a=7.97$ fm. Fig. \ref{fig:strength} therefore includes both cases, i.e. $a=4.70$ fm and 7.97 fm. In this study, we adopt $a=7.97$ fm because one of our goals is to put an upper limit on the resonance strength. }

{One can find in Fig. \ref{fig:strength} that the resonance strength in \citet{bennett} and three of the resonances in \citet{bravo} are excluded, two in \citet{bravo} are marginally consistent with our upper limit, and those in \citet{cooper} are consistent with the upper limit. }

\section{Implication for WD-WD Mergers}
\subsection{Basic Scenarios of WD Binary Mergers}

\begin{figure}
\centering
\includegraphics[width=7cm]{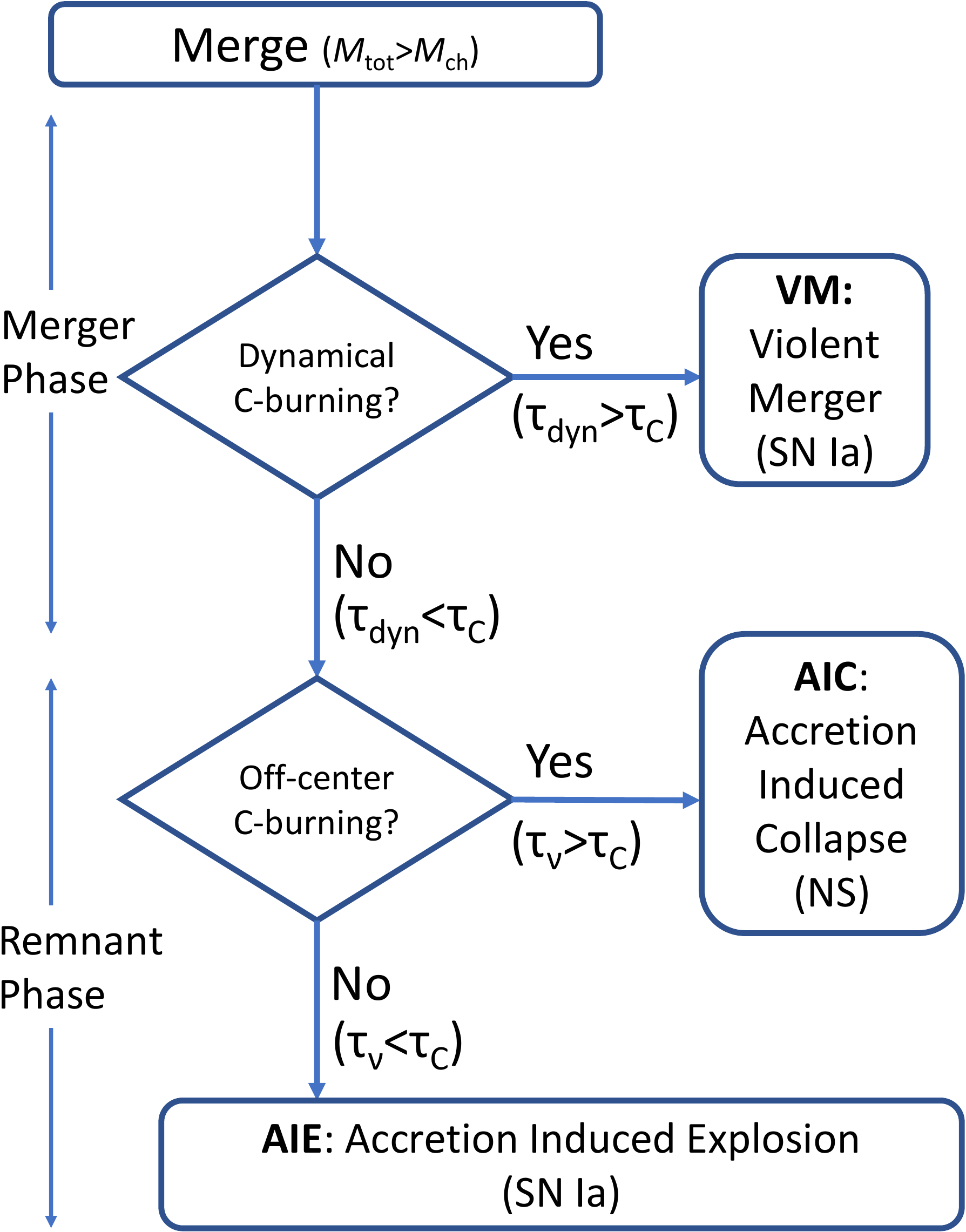}
\caption{The standard scenario on the evolution of WD-WD binary mergers.}
\label{fig:evo}
\end{figure}

The evolution of WD binary mergers into SN Ia explosions or collapse to NSs depends on $\tau_\mathrm{C}$, $\tau_\nu$ and $\tau_\mathrm{dyn}$, where $\tau_\mathrm{dyn}$ is the typical dynamical timescale of the system.
Given the energy generation rate of carbon burning $\epsilon_\mathrm{C}$ \citep{blinnikov} and the neutrino cooling rate $\epsilon_\nu$ \citep{itoh}, the timescales corresponding to these processes are $\tau_\mathrm{C}=C_\mathrm{P}T/\epsilon_\mathrm{C},\;\tau_\nu=C_\mathrm{P}T/\epsilon_\nu$, where $C_\mathrm{P}$ is {the} specific heat. The fate of WD-WD binary mergers is illustrated in Fig. \ref{fig:evo}. 

If the total mass is smaller than the Chandrasekhar mass $M_\mathrm{ch}$, a massive white dwarf remains. Let us assume here that the total mass of the two WDs is larger than $M_\mathrm{ch}$. If the secondary WD accretes gas materials violently onto the primary WD, carbon burning occurs dynamically and detonation propagates throughout the system, which leads to a SN Ia explosion. This occurs when $\tau_\mathrm{dyn}>\tau_\mathrm{C}$ and is called the violent merger \citep[VM;][]{pakmor10,pakmor12}. If the detonation does not occur in the merger phase for $\tau_\mathrm{dyn}<\tau_\mathrm{C}$, a remnant that has a cold core, a hot envelope and an outer disk is formed. If the ignition condition for carbon burning $\epsilon_\mathrm{C}>\epsilon_\nu$ or equivalently $\tau_\mathrm{C}<\tau_\nu$ is satisfied in the envelope, a carbon burning front propagates through the core and converts the CO WD into an ONeMg one. Once the ONeMg WD forms, it cannot support the mass because of electron capture and it collapses into a NS. This scenario is referred to as the accretion induced collapse \citep[AIC;][]{nomoto91}.

For $\tau_\nu<\tau_\mathrm{C}$, the WD will explode as a SN Ia due to the high central density (namely, accretion induced explosion; AIE).
\subsection{Ignition Condition of Carbon Burning}
\begin{figure}
\centering
\includegraphics[width=8.5cm]{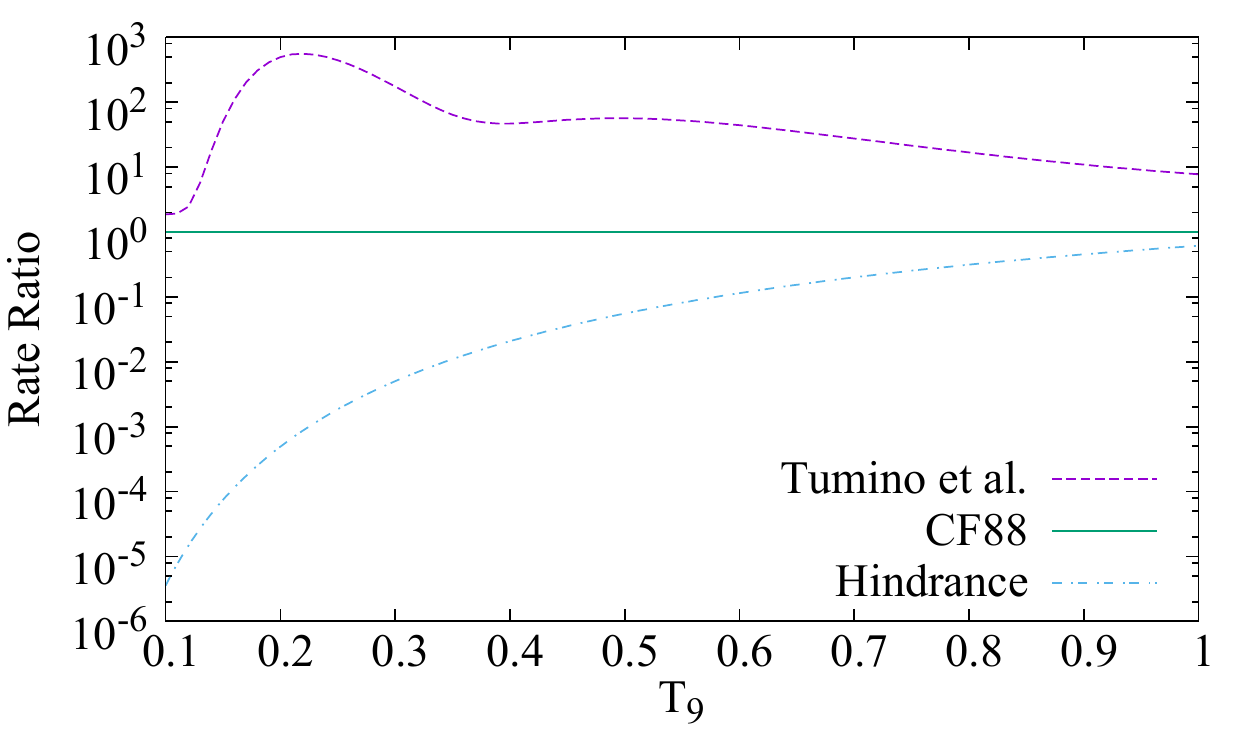}
\caption{The $^{12}$C+$^{12}$C reaction rate normalized by the CF88 rate. }
\label{fig:reacrate}
\end{figure}

\citet{tumino18} succeeded in measuring the cross sections below 2 MeV using the Trojan Horse Method \citep{thm} with the three-body processes $^{12}$C($^{14}$N, $\alpha^{20}$Ne)$^2$H and $^{12}$C($^{14}$N, $p^{23}$Na)$^2$H. Several resonances were found near the Gamow peak resulting in reaction rates roughly 25 times that of  CF88 at $T=5\times10^8$ K. The reaction rates as a function of the temperature are shown in Fig. \ref{fig:reacrate} {(purple broken line)}. On the other hand, \citet{jiang} suggested that the $S^\ast$-factor can decrease in the low-energy region. This ``hindrance'' model is shown in the blue broken line in Fig. \ref{fig:reacrate}.

\begin{figure}
\centering
\includegraphics[width=8.5cm]{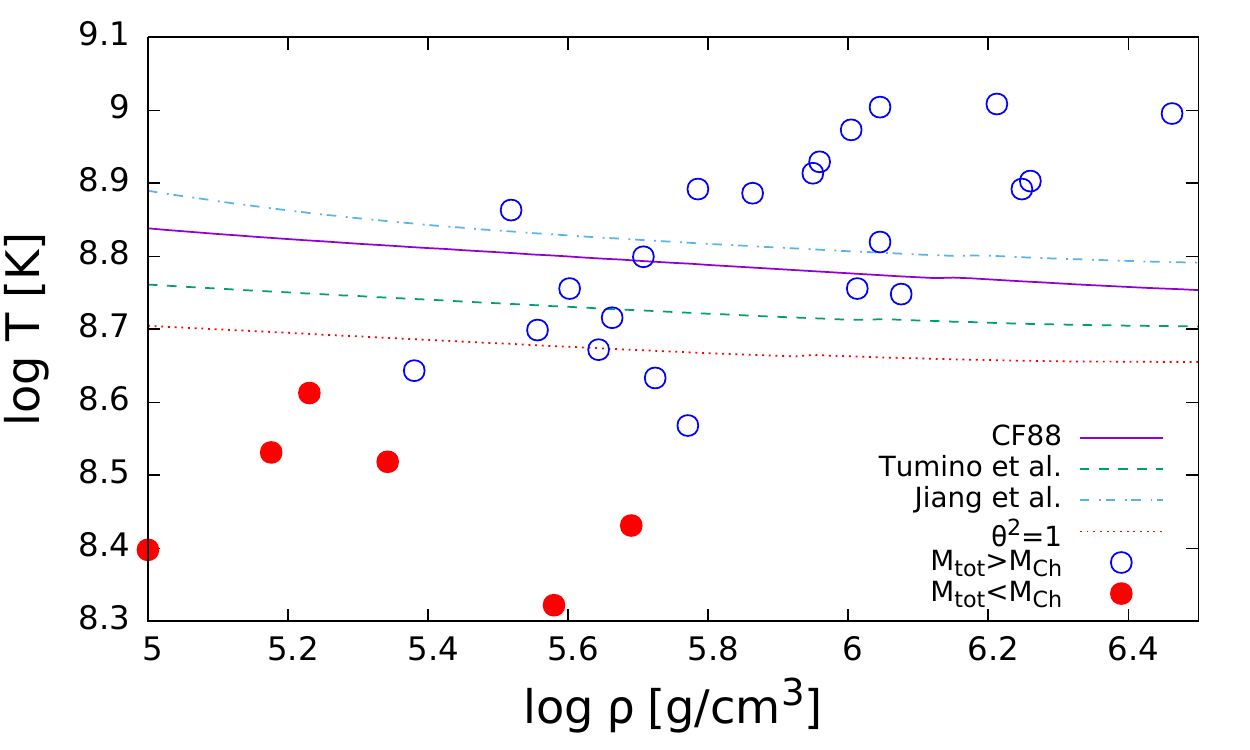}
\caption{The ignition curve $\epsilon_\mathrm{C}=\epsilon_\nu$ (i.e. $\tau_\mathrm{C}=\tau_\nu$) {for the C-burning rate} with and without the resonance contribution. The ignition temperature with the experimental reaction rate \citep{tumino18} is drawn in the green line. The blue line shows the ignition temperature with the hindrance model \citep{jiang}. The circles are calculated results of SPH simulations of \citet{sato15,sato16}. The blue open points are for systems with $M_\mathrm{tot}>1.4M_\odot$ and the red closed points are for systems with $M_\mathrm{tot}<1.4M_\odot$. }
\label{fig:ignition}
\end{figure}

The ignition curve, shown in Fig. \ref{fig:ignition}., is defined to be the temperature and density at which the ignition condition, $\epsilon_\mathrm{C}=\epsilon_\nu$, is satisfied. Here, $\theta^2$ is the dimensionless reduced width \citep[e.g.][]{clayton} of an assumed {Breit-Wigner} resonance at a resonance energy of $E_\mathrm{R}=1.37$ MeV \citep{chiba}, thus the $\theta^2=1$ corresponds to the theoretical lower limit of the ignition temperature. {For this figure, the approximation of a narrow resonance is adopted.} Also shown are the calculated results of smoothed particle hydrodynamics (SPH) simulations of WD mergers  \citep{sato15,sato16}, which
indicate the highest temperature observed in the simulation and the density at that point. The blue open and closed red circles in this figure represent the calculated results from various combinations of progenitor in the SPH simulations for systems of total mass with $M_\mathrm{tot}>1.4M_\odot$ and $M_\mathrm{tot}<1.4M_\odot$, respectively. The systems above the ignition curve will collapse to NSs, while those below the curve will explode as SNe Ia, if the total mass is larger than $M_\mathrm{ch}\sim 1.4M_\odot$. Some of the systems which result in AIE using only the non-resonant CF88 rate {evolve into the AIC eventually since} including the resonance contribution lowers ignition temperature. 

Secondary reactions contribute to energy generation in carbon burning. \citet{blinnikov}, followed by \citet{dan} and \citet{sato15}, suggested an average $Q$-value, $Q_\mathrm{ave}=9.27$ MeV, assuming a branching ratio of 1/2 for the reactions $^{12}$C($^{12}$C, $\alpha$)$^{20}$Ne ($Q=4.6$ MeV) and $^{12}$C($^{12}$C, $\gamma$)$^{24}$Mg ($Q=13.9$ MeV). This assumption, however, should be revised {using} detailed reaction network calculations \citep{chamulak,iliadis}. \citet{iliadis} performed a network calculation with constant temperature and density at $T_9=0.9$ and $\rho=10^5\;\mathrm{g/cm^3}$ with initial composition of $X(^{12}\mathrm{C})=0.25,\;X(^{16}\mathrm{O})=0.73,\;X(^{20}\mathrm{Ne})=0.01$ and $X(^{22}\mathrm{Ne})=0.01$, which is analogous to the environments in the WD mergers. It was shown that the major secondary reactions are $^{23}$Na($p$, $\alpha$)$^{20}$Ne ($Q=2.38$ MeV) and
$^{16}$O($\alpha,\;\gamma$)$^{20}$Ne ($Q=4.73$ MeV), which lead to the net reaction of  $2\times^{12}$C+$^{16}$O$\rightarrow2\times^{20}$Ne ($Q=9.35$ MeV). This $Q$-value 9.35 MeV of the net reaction is so similar to the averaged $Q_\mathrm{ave}$ {from} \citet{blinnikov} that this difference does not practically affect the ignition condition. However, it is noted that the physical situation is different. \citet{chamulak} also performed a network calculation with higher densities $\rho\geq 10^9\;\mathrm{g/cm^3}$. It is desirable to use their effective $Q$-values for realistic burning processes near the center of WDs or the surface of NSs.
\subsection{Fate of WD Binary Mergers}
\begin{figure}
\centering
\includegraphics[width=75mm]{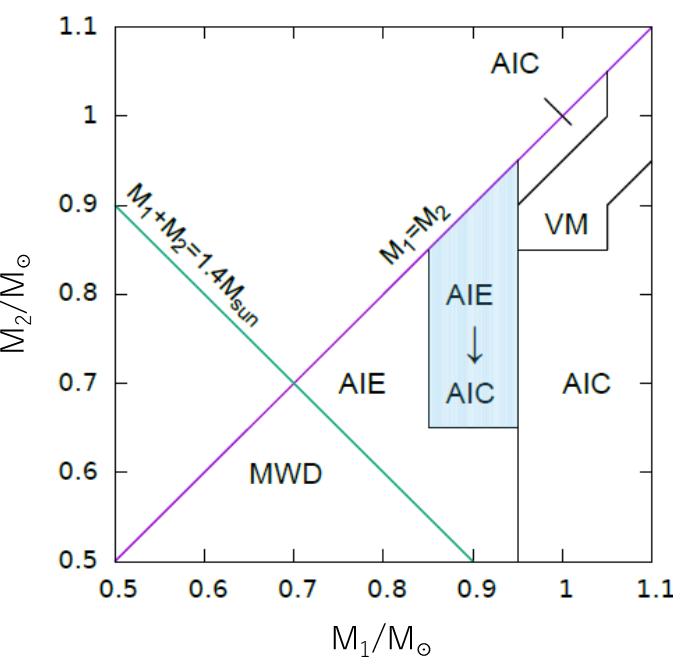}
\caption{The outcome of the WD-WD mergers on the $M_1$-$M_2$ parameter plane with the enhanced reaction rate of \citet{tumino18}. The colored area at $M_1\sim0.9M_\odot$ shows the systems which change their fate from AIE to AIC when the resonance is assumed. MWD is an abbreviation of massive white dwarfs.}
\label{fig:fate}
\end{figure}

Fig. \ref{fig:fate} shows the merger outcomes for the calculated results using the CF88 rate and the enhanced reaction rate by \citet{tumino18}. The masses of the primary and secondary WDs are denoted by $M_1$ and $M_2$, respectively (i.e. $M_1>M_2$). 
The difference between the results for the two rates is the shaded region in which mergers with $M_1=0.9M_\odot$ {go} to the AIC path for the case including the resonance, while they {go} to the AIE path for the non-resonant CF88 rate.

Combining these results with the mass distribution of single DA\footnote{WDs can be classified by their optical spectra. DA is a class with strong hydrogen lines.} WDs extracted from Data Release 4 of the Sloan Digital Sky Survey (SDSS) \citep{kepler}, we can estimate a SN Ia rate which comes from the WD-WD mergers. \citet{badenes} showed that the Galactic event rate of WD-WD mergers per unit stellar mass is estimated to be $1.4^{+3.4}_{-1.0}\times 10^{-13}\;M_\odot^{-1}\mathrm{yr}^{-1}$ from spectroscopic data. The VM rate is $8.4^{+20.4}_{-6.0}\times10^{-17}\;M_\odot^{-1}\mathrm{yr}^{-1}$ and nearly independent of reaction rates because the critical temperature to cause the dynamical instability necessary for a VM is as high as $T_9>1.5$. However, the AIE rate changes depending on {the resonance contribution}. Its rate is $1.3^{+3.2}_{-0.92}\times10^{-14}\;M_\odot^{-1}\mathrm{yr}^{-1}$ for the non-resonant CF88 rate, while it decreases to $1.2^{+2.9}_{-0.85} \times 10^{-14}\;M_\odot^{-1}\mathrm{yr}^{-1}$ for the rate plus resonance contribution, compared to the rate of SNe Ia in Sbc\footnote{Sbc is one of morphological classes of barred spiral galaxies, to which the Milky Way Galaxy is believed to belong.} galaxies with the stellar mass of the Milky Way is $\sim (1.1\pm0.2)\times10^{-13}\;M_\odot^{-1}\mathrm{yr}^{-1}$ \citep{li}. WD-WD mergers account for only $\sim12\%$ of SNe Ia for {the case of} the CF88 rate and the situation is almost the same for the \citet{tumino18} rate. The result is {summarized} in Table \ref{tab:table} for several reaction rates.

Recently, \citet{maoz} estimated the Galactic WD-WD merger rate as $(7\pm 2)\times 10^{-13}\;M_\odot^{-1}\mathrm{yr}^{-1}$  using spectroscopic data from the ESO-VLT SN Ia Progenitor Survey (SPY). This is $\sim 5$ times larger than the  estimate of \citet{badenes}. Event rates calculated from \citet{maoz} are  summarized in Table \ref{tab:table2}. Using \citet{tumino18} reaction rate, the DD scenario would be responsible for $\sim55\%$ of the SNe Ia rate.

The NS birth rate has been estimated to be $10.8^{+7.0}_{-5.0}$ NSs/century, while the CCSN rate is estimated to be $1.9\pm1.1$ SNe/century from measurements of $\gamma$-ray from $^{26}$Al \citep{diehl,keane}, suggesting that the origin of NSs is supplemented by  the AIC path of the WD mergers. The estimated AIC rate is tabulated in Table \ref{tab:table} and \ref{tab:table2}. The \citet{tumino18} reaction rate can increase this rate by {$\sim20\%$}. However, the enhanced AIC rate does not completely solve the birthrate problem of the NSs.

\begin{figure*}
\centering
\includegraphics[width=8cm]{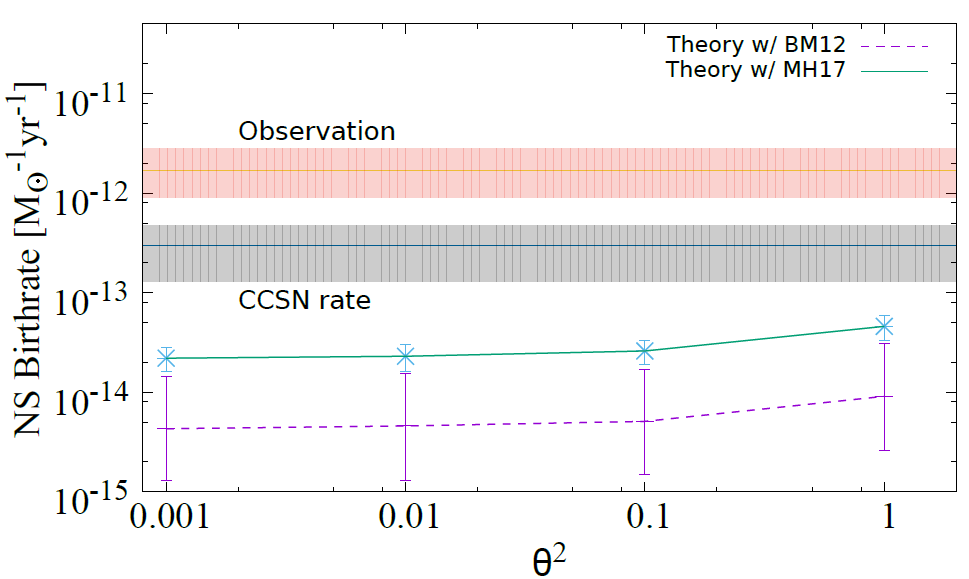}
\includegraphics[width=8cm]{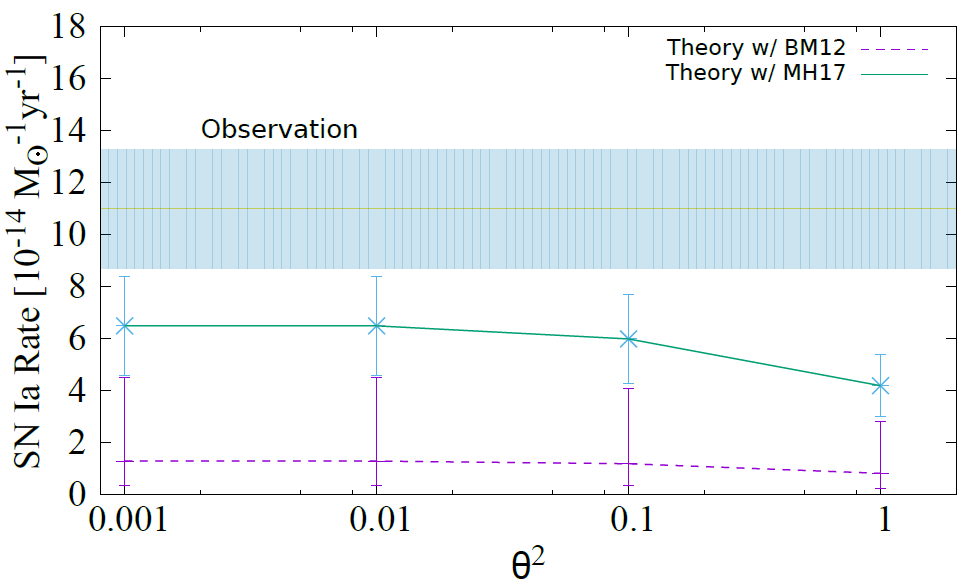}
\caption{(Left) The birthrate of NSs. The bands show observational uncertainties. The points show the theoretical AIC rates with the WD-WD merger rates \citet{badenes} (BM12) and \citet{maoz} (MH17). (Right) The event rate of SNe Ia. The band show observational uncertainty. }
\label{fig:aic}
\end{figure*}

The astrophysical event rates are summarized in Fig. \ref{fig:aic} as a function of $\theta^2$. In this figure, the resultant event rates with the \citet{tumino18} reaction rate correspond to $\theta^2\approx0.1$ . 

\begin{table*}
\centering
\caption{The event rates of the AIC, AIE and VM paths in units of $10^{-14}\;/M_\odot/\mathrm{yr}$ and the ratio of NSs and SNe Ia that can be explained in each scenario. The Galactic WD-WD rate is from \citet{badenes}. \label{tab:table}}
 \begin{tabular}{c|ccc|cc}
& AIC & AIE & VM & AIC/NS & (VM+AIE)/SNIa\\\hline
CF88 & $0.43^{+1.0}_{-0.30}$ & $1.3^{+3.2}_{-0.92}$ & $0.0084^{+0.0204}_{-0.0060}$ & $0.0025^{+0.013}_{-0.0020}$ & $0.12^{+0.40}_{-0.091}$ \\
\citet{tumino18} & $0.51^{+1.2}_{-0.36}$ & $1.2^{+2.9}_{-0.85}$ & $0.0084^{+0.0204}_{-0.0060}$ & $0.0030^{+0.016}_{-0.0025}$ &$ 0.11^{+0.36}_{-0.084}$ \\
$\theta^2=1$ & $0.91^{+2.2}_{-0.65}$& $0.83^{+2.0}_{-0.59}$ & $0.0084^{+0.0204}_{-0.0060}$ & $0.0054^{+0.029}_{-0.0045}$ & $0.076^{+0.25}_{-0.058}$ \\
Hindrance&$0.36^{+0.87}_{-0.26}$&$1.4^{+3.4}_{-1.0}$&$0.0084^{+0.0204}_{-0.0060}$&$0.0021^{+0.011}_{-0.0017}$&$0.13^{+0.42}_{-0.10}$\\
 \end{tabular}
\end{table*}
\begin{table*}
\centering
\caption{Same as Table \ref{tab:table}, but \citet{maoz} is adopted as the WD-WD merger rate. \label{tab:table2}}
\begin{tabular}{c|ccc|cc}
 & AIC & AIE & VM & AIC/NS & (VM+AIE)/SNIa\\\hline
CF88 & $2.2\pm0.6$ & $6.5\pm1.9$ & $0.042\pm0.012$ & $0.013^{+0.018}_{-0.007}$ & $0.60^{+0.37}_{-0.25}$ \\
\citet{tumino18}& $2.6\pm0.7$ & $6.0\pm1.7$ & $0.042\pm0.012$ & $0.015^{+0.021}_{-0.008}$ & $0.55^{+0.34}_{-0.23}$ \\
$\theta^2=1$ & $4.6\pm1.3$& $4.2\pm1.2$ & $0.042\pm0.012$ & $0.027^{+0.038
}_{-0.015}$ & $0.38^{+0.24}_{-0.15}$ \\
Hindrance&$1.8\pm0.5$&$7.0\pm2.0$&$0.042\pm0.012$&$0.011_{-0.006}^{+0.013}$&$0.64^{+0.39}_{-0.27}$\\
\end{tabular}
\end{table*}

{The ignition temperature of carbon burning increases if the hindrance model \citep{jiang} is adopted, as shown in Fig. \ref{fig:ignition}. This leads to a higher AIE rate of $1.4^{+3.4}_{-1.0}\times10^{-14}/M_\odot/\mathrm{yr}$ and a lower AIC rate of $3.6^{+8.7}_{-2.6}\times10^{-15}/M_\odot/\mathrm{yr}$ than those calculated with CF88, assuming the event rate of mergers of \citet{badenes} (Table \ref{tab:table}). If we use the event rate estimated by \cite{maoz}, these results become $\sim$ 5 times larger (Table \ref{tab:table2}).}

\subsection{Model Uncertainties}
Additional uncertainties are intrinsic to the hydrodynamic models. The SPH simulation in \citet{sato15} stops its calculations at the end of the early remnant phase. {Subsequent evolution is dominated by physical viscosity, which has not been treated,} despite the fact that the carbon burning can start in the viscous evolution phase \citep{shen,schwab}. Therefore, some of the systems which go to the AIE path in this study may change their fate to the AIC path. Realistic simulations of mergers with viscosity are desirable to acquire the conclusive result.

\citet{sato15} checked the convergence of their results by changing the numerical resolution, reporting that the maximum temperature nearly converges in the remnant phase, while it gradually increases with higher resolutions in the merger phase. Hence the fate of some systems may change from AIC to VM if simulations for higher-resolution studies are carried out.
\section{Summary and Future Prospect} \label{sec:discussion}
The  low energy resonances in the $^{12}$C+$^{12}$C fusion reaction were studied. Resonant reaction rates were applied to WD-WD binary mergers. The enhanced reaction rate results in a lower ignition temperature, leading to a higher probability of finding WD-WD mergers reaching the AIC. {This could increase the birthrate of the Galactic NSs making} the fraction of the WD mergers in the progenitors of SNe Ia smaller{. Although this {result} favors a partial solution of the NS birthrate problem,} the contribution of the DD scenario to SNe Ia is still largely subject to observational errors.

The result by \citet{tumino18} significantly impacts a wide range of astrophysics, though the validity of this method is a subject of debate. \citet{akram} pointed out that the Coulomb interaction is so large that the plane-wave approach used in \citet{tumino18} could be questionable at the energies used in their experiment and that the $R$-matrix analysis  should be reevaluated to account for identical bosons, etc. {\citet{tumino18b} quickly replied counter discussion against \citet{akram}}, but theoretical assumptions used in the data analysis are to be carefully studied. Therefore, both measurements of the low energy cross sections and theoretical analysis of the molecular resonances in the $^{24}$Mg nuclear system are highly desirable to confirm the existence of the resonances and to determine the resonance parameters $E_\mathrm{R}$, $\Gamma_\mathrm{tot}${, and partial decay widths, especially of the carbon channel $\Gamma_\mathrm{C}$}. The Laboratory for Underground Nuclear Astrophysics \citep[LUNA;][]{luna} is planning to measure the low energy cross sections down to the Gamow peak.

\section*{acknowledgements}
The authors thank Shigeru Kubono and Izumi Hachisu for helpful discussions. KM was supported by a grant from the Hayakawa Satio Fund awarded by the Astronomical Society of Japan. {MAF was supported by the National Science Foundation under Grant No. PHY1712382.} This work is supported by JSPS KAKENHI Grant Numbers JP15H03665 and JP17K05459.

%
%

\end{document}